\documentstyle[preprint,eqsecnum,aps]{revtex}

\begin{document}
\draft
\title{ Fluid  accretion
onto a spherical  black hole: relativistic  description versus
Bondi model}
\author{Edward Malec}
\address{Jagellonian University,
Institute of Physics, 30-059 Krak\'ow, Reymonta 4,  Poland}
\date{\today}
\maketitle
\begin{abstract}

 We describe general-relativistically a spherically symmetric
 stationary fluid accretion onto a black hole.
Relativistic effects enhance mass accretion, in comparison to
the Bondi model predictions, in the case when
backreaction is neglected. That enhancement
 depends on the adiabatic index and
the asymptotic gas temperature and it can magnify accretion by one order
in the ultrarelativistic regime.

\end{abstract}

\centerline{PACS: 04.40.-b, 98.62.Mw, 04.20.-q}

\section{ Introduction.}

In this paper we re-examine the spherical gas accretion onto a black hole,
 paralleling previous studies of
 fluid accretion of Michel and  Shapiro-Teukolsky
  (\cite{Michel}, \cite{Shapiro}). It is  shown that relativistic
 effects can lead to a bigger  mass accretion than that
predicted by the corresponding Bondi model \cite{Bondi}.

The order of this paper is as follows. Next section presents  spherically
symmetric Einstein equations expressed in the language of extrinsic
curvatures. A suitable choice of a gauge condition
leads to a "comoving coordinates"
\cite{Tolman} formulation that is particularly
suitable for the description of selfgravitating fluids.
In Section III we show that the original set of integro-differential
equations can be reduced to an integro-algebraic problem,
 whose solution would constitute
a new stationary, general-relativistic solution
of selfgravitating polytropic
 fluids. That model is complete -  it includes the
back effect  exerted by matter onto a metric -  therefore
it is capable to describe    a stationary phase of the
interaction of (even) heavy clouds of gas with a relatively light center.
 Section IV discusses a case when backreaction
can be neglected. Under some circumstances, an accretion
is described by a set of purely algebraic equations.
Section V proves several quantitative and qualitative properties of accreting
solutions. It
is shown that the  Bondi model relation between the asymptotic and sonic
speeds of sound appears as a limiting case of relativistic
formulae.
  Section VI compares predictions
of the Bondi model and of the relativitic solution
without backreaction.
 Relativistic magnification  of the mass accretion becomes noticeable in
the case of infall of a hot gas, when the correction factor can be bigger
than $2.4(1+{1\over \Gamma })$, where $\Gamma $ is the polytropic index.

\section{Equations.}

We will use a spherically symmetric  line element
\begin{equation}
ds^2= -N^2dt^2+adr^2+R^2\bigl( d\theta^2+\sin^2 \theta \bigr) d\phi^2
\label{i1}
\end{equation}
where $N, a$ and $R$ depend on $t$ (asymptotic time variable) and  a
coordinate radius $r$.
we will work in extrinsic curvature variables.
Thus we need the mean curvature of centered  two-spheres
in a Cauchy slice,
\begin{equation}
p = {2\partial_rR \over \sqrt{a} R}
\label{i2}
\end{equation}
and the extrinsic curvature
\begin{eqnarray}
&&trK={\partial_t(\sqrt{a}R^2) \over N \sqrt{a} R^2}\nonumber\\
&&K_r^r={1\over 2N a}\partial_0a,\nonumber\\
&&K_{\phi  }^{\phi  }=
 K_{\theta   }^{\theta  }={\partial_0R \over N R}={1\over 2}(trK-K_r^r).
\label{i3}
\end{eqnarray}

Let  $T_{\mu }^{\nu }$ be the  energy-momentum tensor of matter fields,
$\rho =-T_0^0$ and $j_r= NT^0_r$, $R_{\mu \nu }$ be the Ricci tensor and
$R$ the Ricci scalar.

The Einstein  constraint equations $R_{0\mu }-g_{0\mu }R/2=8\pi T_{0\mu }$
can be integrated to yield,
assuming asymptotic flatness,

\begin{equation}
Rp=
2\sqrt{1-{2m\over R}+{8\pi \over R}\int_R^{\infty }\tilde R^2 \rho
d\tilde R+ \tau },
\label{i8}
\end{equation}
\begin{equation}
RK_r^r-RtrK={C_1-8\pi \int_0^R{2\tilde R^2j_r\over \sqrt{a}p}d\tilde R\over
R^2}-2{\int_0^RtrK \tilde R^2d\tilde R\over R^2},
\label{i10}
\end{equation}
where  m is the asymptotic  mass and
\begin{equation}
\tau ={3\over 4R}\int_R^{\infty } \tilde R^2(K_r^r)^2d\tilde R  -{1\over 4R}
\int_R^{\infty }\tilde R^2(trK)^2d\tilde R-{1\over 2R}\int_R^{\infty } trK
K_r^r\tilde R^2 d\tilde R.
\label{i9}
\end{equation}
Impose the integral gauge condition
\begin{equation}
\tau = \bigl( {R(trK-K_r^r)\over 2}\bigr)^2
\label{f1}
\end{equation}
where $\tau $ is given in (\ref{i9}).
One can show that in vacuum (\ref{f1}) is satisfied identically.

Differentiation of (\ref{f1}) with respect $r$ yields, after some algebra,
\begin{equation}
R(trK-K_r^r) {16\pi j_rR\over p}=0
\label{f2}
\end{equation}
which implies
\begin{equation}
j_r=0=U_i
\label{f3}
\end{equation}
in  geometries without minimal surfaces and with $tr K\ne K_r^r$.
Thus in this gauge coordinates are  "comoving" - each particle of matter
carries   a fixed value of a radial coordinate "$r$".

 The energy-momentum tensor of a  selfgravitating fluid  reads,
 in comoving coordinates,
\begin{equation}
T_{\mu \nu }=( \rho +\tilde p )U_{\mu }U_{\nu }+\tilde p g_{\mu \nu }.
\label{f0}
\end{equation}
Here   $U_{\mu }U^{\mu }=-1$.  Notice that  the presure  is
  $\tilde p=T_r^r=T_{\theta }^{\theta }$.
This space-time foliation is regular even at the vicinity of the boundary of
a black hole, in  contrast  with other approaches \cite{Michel},
\cite{Shapiro} in  which  the Schwarzschild geometry is foliated 
by polar gauge slices.

Define a mass function
\begin{equation}
m(R(r))=2\pi \int_0^r\tilde R^3p  \rho dx
\label{f5}
\end{equation}
where $\tilde R $ is an areal radius.
The mass  evolves as follows
\begin{equation}
\partial_0 m(R(r)) =
-2\pi [N R^3(tr K- K_r^r)\tilde p](r)
\label{f6}
\end{equation}
 Direct differentiation of $m(R)$ gives
\begin{equation}
{\partial_rm(R)\over \sqrt{a}}= 2\pi R^3p\rho .
\label{f7}
\end{equation}
 The remaining relevant equations are the
two continuity equations
\begin{equation}
N \partial_r \tilde p
 +  \partial_rN (\tilde p+\rho)=0,
 \label{f8}
 \end{equation}
\begin{equation}
\partial_0  \rho=
-N tr K (\tilde p +\rho  ).
 \label{f9}
\end{equation}
and the Einstein evolution equation

\begin{equation}
\partial_t(K_r^r-trK)= {3N \over 4}(K_r^r)^2
-{N p^2\over 4}-{p\over \sqrt{a}} \partial_rN +{N  \over R^2}
+8\pi N T_r^r +{3\over 4}N (trK)^2-{3N \over 2}
trK K_r^r
\label{i7}
\end{equation}
The rate of accretion $\dot m$ of mass along orbits of a constant areal radius
$R$ is equal to
\begin{equation}
\dot m(R) \equiv (\partial_0-\dot R\partial_R) m(R(r)) =
-4\pi N R^2U(\tilde p+\rho ).
\label{f10}
\end{equation}
where
\begin{equation}
U\equiv \partial_0R/N={R\over 2}(trK -K_r^r).
\label{f11}
\end{equation}

\section{Stationary description of a selfgravitating fluid.}

All results of  this section
hold true   for systems with collapsing or  exploding matter.
  Assume a  compact cloud of a fluid.
We will say that an  accretion (explosion) is stationary if

i) the mass accretion
$$\dot m\equiv (\partial_0-\dot R\partial_R) m(R(r))|_{R=const}$$
on a central body  is constant in time;

ii) the  radial fluid velocity $U={\partial_0R\over N}$ is
constant at a fixed
value of the areal radius $R$, $(\partial_0-\dot R \partial_R )U=0$;

iii) the  energy density at a fixed areal radius does not change in time;

iv) asymptotically, i. e., close to the outer boundary of the collapsing
 (exploding) gas, its  speed  $U$  is much smaller than the speed of
sound $a^2=\partial_{\rho }\tilde p$, $U_{\infty }\approx 0$.
(Expanding fluid, in turn, would be
subject to a condition $U\approx 0$ at the inner boundary
\cite{Michel}.)

At first,  we shall prove   the following     fact

{\bf Theorem 1.}  Under conditions  i-iii),
$\dot m$  does not depend on
$R$ within the fluid filled  zone,
$\partial_R\dot m=0$.

{\bf Proof. }

Equations (\ref{i10}) and (\ref{f11}) yield
\begin{equation}
\partial_R(UR^2)= R^2trK.
\label{4.1}
\end{equation}
>From $\dot m(R)= -4\pi NUR^2(\rho +\tilde p)$
we obtain $\partial_R\dot m=I +II +III$,  where
\begin{eqnarray}
&&I=N(\rho +\tilde p) \partial_R(UR^2)\nonumber\\
&&II=NUR^2\partial_R\tilde p +UR^2(\tilde p+\rho )\partial_RN\nonumber\\
&&III=NUR^2\partial_R\rho .
\label{4.2}
\end{eqnarray}
Using (\ref{4.1}) one writes $I=NR^2 trK (\tilde p +\rho )$, while
the stationarity condition ii) allows one to write $III=R^2\partial_0\rho=
-NR^2trK (\tilde p +\rho )$ (the second equality follows from  (\ref{f9})).
Thus $I+III=0$; since $II=0$ (due to the momentum conservation
(\ref{f8})), we arrive at  $\partial_R\dot m(R)=0$.

Assume the equation of  state
\begin{equation}
\tilde p = K\rho^{\Gamma },
\label{4.3}
\end{equation}
$\Gamma $ being a constant and define the speed of sound
 as $a^2=\partial_{\rho } \tilde p$.
We assume that $1\le \Gamma \le 5/3$, since we are primarily interested in
 comparing predictions with the Bondi model, but it is quite
 likely that much of the forthcoming analysis
 applies to adiabatic indices in the standard in astrophysics range (1, 2).
 
 Let us point out that astrophysicists
\cite{Shapiro} use  a different equation of state,
 $\tilde p = Cn^{\Gamma }$ (where  $n$ is the
baryon number density); that reads in our notation
\begin{equation}
\tilde p = C\times \exp (\Gamma \int d\rho {1\over \rho +K\rho^{\Gamma }}).
\end{equation}
Both approaches agree for $\Gamma =1$ or in
 the newtonian limit when $\Gamma \ne 5/3$, but they disagree for
$\Gamma =5/3$.
The  momentum conservation equation (\ref{f8}) can be integrated,
\begin{equation}
a^2 = -\Gamma +{\Gamma +a^2_{\infty }\over N^{\kappa }},
\label{4.4}
\end{equation}
where $\kappa ={\Gamma -1\over \Gamma } $ and the integration constant
$a^2_{\infty }$ is equal to the asymptotic speed of sound of a fluid.

(\ref{4.4}) asymptotically ($m/R<<1$) yields the Bernoulli
equation, hence it can be regarded
as the general-relativistic version  of the  latter.

>From the relation between  pressure and energy density, one
obtains, using   equation (\ref{4.4})
\begin{equation}
\rho = \rho_{\infty }\Bigl( a/a_{\infty }\Bigr)^{2/(\Gamma -1)}
=\rho_{\infty } \bigl[ -{\Gamma \over a_{\infty }^2}  +
{{\Gamma \over a_{\infty }^2}  +1\over N^{\kappa  }}\bigr]
^{1\over \Gamma -1},
\label{4.17}
\end{equation}
where the constant  $\rho_{\infty }$ is equal to the asymptotic mass density
of a   fluid.
>From the evolution equation  one gets, using the stationarity assumption,
 \begin{eqnarray}
\dot R \partial_R U  =
   {1\over 4}(pR)^2\partial_RN- m(R){N\over R^2}-4\pi RN\tilde p.
 \label{4.5}
\end{eqnarray}

Equations (\ref{i8}) and (\ref{i10}) give
\begin{equation}
U\partial_RU= {pR\over 4}\partial_R(pR)-{m(R)\over R^2}+4\pi \rho R.
\label{4.7}
\end{equation}
Comparison of (\ref{4.7})   with (\ref{4.5}) yields an ordinary
differential equation
\begin{equation}
\partial_R\ln ({N\over pR})={16\pi \over p^2R}(\rho +\tilde p);
\label{4.8}
\end{equation}
integration of that, with  the asymptotic condition
at spatial infinity $N={pR\over 2}=1$ leads to the following relation
between the lapse $N$ and the mean curvature $pR$
\begin{equation}
N={pR\over 2} \beta (R)
\label{4.9}
\end{equation}
where
\begin{eqnarray}
\beta (r) = e^{  \int_r^{\infty }16\pi (-\tilde p-\rho )
{1\over p^2s}ds}.
\label{4.10}
\end{eqnarray}
The substitution of $trK$ (as calculated from
the continuity equation  (\ref{f9})) into (\ref{4.1})  gives,
employing the stationarity condition,
\begin{equation}
\partial_R\ln (|U|R^2)=-{\partial_R\rho \over \tilde p +\rho }.
\label{4.11}
\end{equation}
Notice that    $-{\partial_R\rho \over \tilde p +\rho }=
-{\partial_R(\rho +\tilde p) \over \tilde p +\rho }+
{\partial_R\tilde p\over \tilde p +\rho }$. The last term
can be presented in another  form
(due to  relations between $a^2, \rho $ and
$\tilde p$),  ${\partial_R\tilde p\over \tilde p +\rho }=
{\Gamma \over \Gamma -1}\partial_R\ln (a^2/\Gamma +1)$.
That leads to the following solution of (\ref{4.11})
\begin{equation}
U=C{ (a^2/\Gamma +1)^{1/(\Gamma -1)}\over R^2 a^{2/(\Gamma -1)}}.
\label{4.12}
\end{equation}
 The whole set of  equations describing the collapsing stationary
fluid is given by (\ref{4.12}) and the previously written
equations (\ref{4.4}, \ref{4.9}) and (\ref{4.10}).
Calculation of  $\partial_R\ln (a^2+\Gamma ) $, with  $a^2$ given
by (\ref{4.4}) and $N$ being specified above, yields
\begin{equation}
\partial_R\ln (a^2+\Gamma ) = {-4\kappa \over p^2R^3}\Bigl( {m(R)\over R}
+ 4 \pi R^2\tilde p -2U^2+
{1\over 2R^3}\partial_R(U^2R^4) \Bigr) .
\label{4.13}
\end{equation}
One easily obtains  from (\ref{4.12})   that
\begin{equation}
{1\over R^4}\partial_R(U^2R^4)=-{2U^2\over \kappa a^2}
 \partial_R\ln (a^2+\Gamma ).
\label{4.14}
\end{equation}
The insertion  of (\ref{4.14}) into (\ref{4.13}) gives
\begin{equation}
\partial_R(U^2R^4) \Bigl( 1-{4U^2\over a^2p^2R^2}\Bigr) =
{16R\over a^2p^2} ({m(R)\over 2R }+2 \pi R^2\tilde p  - U^2).
\label{4.14a}
\end{equation}
We  define sonic points as such where the the equality $U={pR\over 2}a$
holds true. Let $R_* $  be a radius of a  sonic  point.
Equation (\ref{4.14a}) yields the relation
\begin{equation}
a^2_*(1-{3m_*\over 2R_*}+c_*)=U^2_*= {m_*\over 2R_*}+c_*,
\label{4.15}
\end{equation}
where  $c_* = 2 \pi R^2_*\tilde p_* $, $a^2_*=a^2(R_*)$, $m_*=m(R_*)$
and $U^2_*=U^2(R_*)$.

 The constant $C$ in formula (\ref{4.12})
 can be expressed in terms of $a_*$, $ U_*$ and $R_*$, that
 is as a function  of $c_*$, $m(R_*)$  and  $R_*$. The infall
 velocity $U$ reads
\begin{equation}
U=U_*{R^2_*\over R^2}
\Bigl(  {1 +{\Gamma \over  a^{2}}\over 1+{\Gamma  \over a^2_*} }
\Bigr)^{1/(\Gamma -1)}
\label{4.16}
\end{equation}
Above $U_*$ means a negative square root in the case of falloff towards
a gravity centre and a positive square root in the case of exploding gas.

The rate of accretion of mass (\ref{f10})
 can be conveniently expressed  by characteristics of
the sonic point $R_*$,
\begin{equation}
\dot m =-4\pi R^2_*\rho_{\infty }\Bigl( 1-{3m_*\over 2R_*}+c_*\Bigr)
^{1/2}U_* \Bigl( {a(R_*)\over a_{\infty }}
\Bigr)^{2/(\Gamma -1)} (1+{a^2_{* }\over \Gamma })\beta (R)
\label{4.18}
\end{equation}
For the  sake of completeness we write down the space-time line element
with the areal radius chosen  as the radial coordinate,
\begin{equation}
ds^2= dt^2(-N^2+{4N^2U^2\over (pR)^2}) -4\beta {U \over pR}dtdR
+{4\over (pR)^2}dR^2 +R^2d\Omega^2.
\label{4.19}
\end{equation}
\section{Relativistic accretion: neglecting backreaction.}

The quasi-stationary accretion shall apply to the description of black
holes interacting with a fluid. The description of the accretion onto
other compact bodies (say, neutron stars) is more complex, since there can
appear shocks, that are excluded in our picture. The above model can be valid
only if shocks are absent, for instance when the inner boundary of a
collapsing shell of gas is disconnected from the surface of a compact body.

All hitherto proven results are exact and - under the preceding reservation -
they refer to a fully nonlinear stationary system consisting of a central
mass  and a cloud of gas that would dynamically influence a geometry through
a backreaction.  If the  gas is heavy, comparing with
the central mass, then $\beta (R)$ is nonconstant; metric functions do
depend on the infalling matter.  That means that backreaction should be taken
into account in description of such a  system.

If the contribution of a fluid to the total asymptotic mass
of a system is negligible, i. e.,
\begin{equation}
m_f\equiv \int_{R>2m}dr r^2\rho << m
\label{V.0}
\end{equation}
and $\tilde p \le  \rho $, then   $\beta \approx 1$ and
\begin{equation}
N\approx {pR\over 2}= \approx
\sqrt{1-{2m\over R} +U^2}.
\label{V.1}
\end{equation}
That would suggest that in this case the standard schwarzschildean metric
constitutes a valid approximation. There is, however, one subtle point.
The reasoning of the former section shows that in order to neglect the effect
of backreaction the following condition
\begin{equation}
c_*= 2 \pi R^2\tilde p<<{2m_*\over R}.
\label{V.2}
\end{equation}
must hold at a sonic point.
That can be interpreted as the demand that
not only $N$ is close to $pR/2$ but also
$\partial_RN$ shall be  approximated by $\partial_R(pR)/2$.

We will say that backreaction is negligible if  both conditions (\ref{V.0})
and (\ref{V.2}) hold true.
In such a case $m\approx m_*$ and one obtains  at the sonic point
\begin{equation}
a^2_*(1-{3m\over 2R_*})=U^2_*= {m\over 2R_*},
\label{V.3}
\end{equation}
The remaining two equations describing accretion are
\begin{equation}
U^2={R^{3}_*m\over 2R^4}
\Bigl( { 1\over 1+\Gamma /a^2_* }\Bigr)^{2/(\Gamma -1})
\Bigl(  { 1+{\Gamma \over  a^2}}\Bigr)^{2/(\Gamma -1)}
\label{V.4}
\end{equation}
and
\begin{equation}
a^2 = -\Gamma +{\Gamma +a^2_{\infty }\over N^{\kappa }}.
\label{V.5}
\end{equation}
This is a purely algebraic system
of equations,  describing the fluid accretion in a
fixed space-time (Schwarzschild) geometry.

\section{Relativistic accretion without backreaction.}

Numerical analysis demonstrates - as pointed first by Michel
\cite{Michel} - the existence of two branches of solutions
of the relativistic fluid equations.  An analytic proof is given below.

In the first part we  prove the existence of a sonic point in a black hole
spacetime endowed with a Schwarzschild metric.
That   black hole - fluid system is shown to possess a sonic point;
that leads, through a  construction outlayed in
the second step, to the existence of two   accreting
solutions.

\subsection{Sonic points}

Define
\begin{eqnarray}
&&L=a^2+\Gamma   \nonumber\\
&& P= {a^2_{\infty }+\Gamma \over \Bigl[ 1- {2m\over R}
+U^2\Bigr]^{(\Gamma
-1)/(2\Gamma ) }}
\label{VII.2}
\end{eqnarray}
where   $U^2$ is given by (\ref{V.4}) with parameters $a_*$ and $U^2_*$
specified by (\ref{V.3}).

The equation $L(R_*)=P(R_*)$ for a sonic point can be written as
 $1+y(3\Gamma -1)=3(a^2_{\infty } +
\Gamma )y^{(\Gamma +1)/(2\Gamma )}$,
 where $y=1-3m/(2R_*)$.
One has to demand that $y>0$ (i. e., $R_*>3m/2$),
since at $y=0$ (or $R_*=3m/2$) the
coordinate system breaks down.
Notice a numerical mistake in  \cite{Michel} which led Michel
 to the wrong claim
that sonic points must exist outside a sphere of a radius $6m$. In fact
they may exist even inside a black hole, although - as we point out below
- that would contradict established views on properties of matter.

The left hand of the equation in question is bigger than
its right hand side at $y=0$  while at $y=1$ the opposite holds true.
Since both sides are continuous in $x$, their graphs must intersect
somewhere. Since $1+y(3\Gamma -1) $  increases at a lower rate than
$3(a^2_{\infty } +\Gamma )y^{(\Gamma +1)/(2\Gamma )}$ for $\Gamma \le 5/3$,
there exists a unique sonic point characterised by
$y_*$. The case with $y_*<1/2$ (i. e.,
when $R_*<2m$) is physically noninteresting. In that case the speed
of sound would be bigger than the velocity of light and the dominant energy
condition \cite{hawking} would be broken, even outside of a black hole.
One easily infers that $y_*$ is a monotonously decreasing  function  of the
asymptotic sound density $a^2_{\infty }$. Therefore there exists a critical
value of $a^2_{\infty }$  which separates solutions that are subluminal
from unphysical solutions that become superluminal.

An interesting feature of the Bondi model is  the simple relation
$a^2 _*/a^2_{\infty }={2/(5-3\Gamma )}$ for $\Gamma <5/3$.
Below   we will show that  this relation appears
 in the nonrelativistic limit of  a relativistic formula.

{\bf Theorem 2}. Let  $a^2_{\infty }$ and $a^2_*$ be the
asymptotic and sonic  speeds of sound, respectively.
Define   $\alpha =(\Gamma -1)/(2\Gamma )$
$$A= \alpha \bigl( (9\Gamma -5)\ln 4-1.5\bigl) $$
and
$$B={3\over 2 }\alpha^2 \Gamma (9\Gamma -7).$$
If sonic points are exterior to a black hole then
\begin{equation}
{1\over {5-3\Gamma \over 2}+B{a^2_*\over  1+3a^2_*}}
   \ge { a^2_*\over a^2_{\infty }} \ge  {1\over   {5-3\Gamma \over 2}
   +A{a^2_*\over  1+3a^2_*}}.
\label{VII.3}
\end{equation}
{\bf Proof.}

Define  $x\equiv {m\over 2R_*}$ (or alternatively,
$x=a_*^2/(1+3a_*^2)$), and
\begin{eqnarray}
&&F\equiv   x(1-3x)^{\alpha } +\Gamma (1-3x)^{1+\alpha } -\Gamma (1-3x)
-{5-3\Gamma\over 2}x -Bx^2
\nonumber\\
&& \Psi =x(1-3x)^{\alpha } +\Gamma (1-3x)^{1+\alpha } -\Gamma (1-3x)
-{5-3\Gamma\over 2}x - A  x^2.
\label{VII.4}
\end{eqnarray}
The  sonic point equation can be written  as
\begin{eqnarray}
&&{a^2_{\infty }\over a^2_{* }} - {5-3\Gamma \over 2}  =
  B x  +{F\over x}=\nonumber \\
&&    A x   + {\Psi \over x}.
\label{VII.5}
\end{eqnarray}
It suffices to show that $F\ge 0$ and $\Psi \le 0$.  We shall  deal
with the first inequality. The second derivative of $F$
with respect $x$ reads
\begin{equation}
F''= 3{\Gamma -1\over 2\Gamma }(1-3x)^{\alpha -2}G(x),
\end{equation}
where
\begin{equation}
G(x)={9\Gamma -7\over 2}(1-3x) -3x {\Gamma +1\over 2\Gamma }-
{1 \over 2}(\Gamma -1) (9\Gamma -7) (1-3x)^{2-\alpha }.
\end{equation}
One shows that $G'\le (3\Gamma -5)/\Gamma \le 0 $; thus
  $G(x)$ is  decreasing for $0\le x\le 1/4$ and
$1\le \Gamma \le 5/3$. Therefore if $F''(x_0)=0$ then $F''(x)<0$ for any
$x>x_0$. That means, taking into account that $F''(0)>0$ and $F'(0)=0$,
that if $F'$ vanishes at a point $x_1$, then it must be negative
in the interval $(x_1, 1/4)$.  In conclusion, either $F$ is increasing
(and then it achieves its minimum at $x=0$) or it has a single
extremum (a maximum) in $(0, 1/4)$. Notice now that $F(0)=0$.
 Thence in order to show that $F(x)\ge 0$
 it is enough to show that $F(x)$ is nonnegative at  $x=1/4$, when
\begin{equation}
F(1/4)= (\Gamma +1){1\over 4^{1+\alpha }}+{\Gamma \over 8}- {5\over 8} 
-{3 \over 128\Gamma } (\Gamma -1)^2(9\Gamma -7).
\end{equation}
A numerical calculation shows that $F(1/4)\ge 0$ and the equality is
achieved only at $\Gamma =1$.

In a similar vein one proves the other inequality $\Psi \le0$. At $x=0$
one has $\Psi =0$. On the other hand,
\begin{equation}
\Psi '= {9\Gamma -5\over 2}\bigl (1-(1-3x)^{\alpha }\bigl) -3x\alpha
(1-3x)^{\alpha -1} -
2\alpha  x \Bigl( (9\Gamma -5)\ln 4-{3\over 2}\Bigr)  .
\end{equation}
   Employing the estimate
\begin{equation}
1-(1-3x)^{\alpha }\le 4 \alpha x\ln 4
\end{equation}
which is valid for $0\le x\le 1/4$ and $0.2\ge \alpha \ge 0$, one arrives at
 $\Psi ' \le 0$. Thus the function $\Psi $ is nonnegative, as desired.
That ends the proof.

Let us remark that
(\ref{VII.3})  can be written as, resolving the biquadratic inequalities,
\begin{eqnarray}
&& {\delta_0-\sqrt{\delta_0^2+4a_{\infty }^2\delta_2}\over 2\delta_2}
\nonumber\\
&&  \ge  a^2_*   \ge  \nonumber\\
&&{\delta_0-\sqrt{\delta_0^2+4a_{\infty }^2\delta_1}\over
2\delta_1}
\label{c1}
\end{eqnarray}
where
\begin{eqnarray}
&&
\delta_0=3a^2_{\infty}-{5-3\Gamma \over 2},\nonumber\\
 &&\delta_1={3\over 2}(5-3\Gamma )+ A
 \nonumber\\
&&\delta_2={3\over 2}(5-3\Gamma )+ B.
\label{c2}
\end{eqnarray}
Asymptotically, i. e., for $x\rightarrow 0$ one obtains
the Bondi equality $a^2 _*/a^2_{\infty }={2/(5-3\Gamma )}$ for
$\Gamma <5/3$. If $\Gamma =5/3$ then the above  gives asymptotically
 $1.12\times a_{\infty }
\ge a^2_*\ge 0.8 a_{\infty }$, in a good agreement with the exact formula
$a^2_*=\sqrt{5/6}a_{\infty }$.
If a sonic point is located at  a horizon of a black hole
(that is, $a_*=1$) then (\ref{VII.3}) (or the above inequalities)
yields $0.79 \ge a_{\infty }\ge 0.5$
for $\Gamma =5/3$. Notice also a rough bound  $a^2_*>1.6\times a^2_{\infty }$
which is valid for any $\Gamma $ and $a^2_{\infty }$; outside
of a black hole $a_*^2\le 1$, therefore  one infers that
 the asymptotic speed of  sound is less than 1.  

Let us point also that equation (\ref{VII.5}) implies that
 asymptotic sonic points
can exist only for models with adiabatic indices $\Gamma < 5/3$.

\subsection{Existence proof}

We show that at least two solutions $(a(R), U(R))$ bifurcate from $R_*$.
Define    $a_{\alpha } $ as a solution of the equation
\begin{equation}
\Bigl( { 1+\Gamma /a_{\alpha^2 } \over 1+\Gamma /a^2_*}\Bigr)^{2/(\Gamma -1)}
=(R/R_*)^{7/2 }.
\label{VII.6}
\end{equation}
>From that and (\ref{V.4}) follows that $U^2=U^2_* \sqrt{R_*/R}$ and
\begin{equation}
a^2_{\alpha}= {\Gamma (R_*/R)^{\beta }\over \delta -
(R_*/R)^{\beta }},
\end{equation}
where $\beta = {7\over 4}(\Gamma -1)$ and $\delta = 1+\Gamma /a^2_*$.

A straightforward calculation gives
\begin{equation}
{d\over dR}\ln L(a_{\alpha }) =-{\beta (R_*/R)^{\beta }\over R
(\delta  - (R_*/R)^{\beta })}
\end{equation}
and
\begin{equation}
{d\over dR}\ln P(a_{\alpha }) = -{2(\Gamma -1) R_*U^2_*
(1-{\sqrt{R/R_*}\over 8 })\over \Gamma R^2
\bigl( 1-{3m\over 2R_*}+U_*^2(3-4{R_*\over R}
+\sqrt{R_*\over R})\bigr) }.
\end{equation}
$L$ and $P$ are equal at $R=R_*$ and they are decreasing
in the vicinity of $R=R_*$. Morever, $\partial_RL=\partial_RP$ at
the sonic point $R_*$.  A careful investigation shows, however,  that
 second  derivatives are both  locally positive   and
\begin{equation}
{d^2\over dR^2}\ln P(a_{\alpha })|_{R=R_*}=
{d^2\over dR^2}\ln L(a_{\alpha })|_{R=R_*}\times {29/14 +7a^2_*/2\over
\beta (1+a^2_*/\Gamma ) +1}.
\label{VII.8}
\end{equation}
One observes that
${d^2\over dR^2}\ln P(a_{\alpha })|_{R=R_*}\ge
{d^2\over dR^2}\ln L(a_{\alpha })|_{R=R_*}$  if $\Gamma < 79/49$.
This reasoning can be  valid for adiabatic indices $\Gamma \le 5/3$ assuming
that the exponent 7/2 in (\ref{VII.6})
  is replaced   by $x\epsilon (-\infty, 4.5-\sqrt{1.5})$ \cite{Rosz}.
Therefore  $\partial_RL<\partial_RP$,
for $R > R_*$ and $\partial_RL>\partial_RP$ for $R < R_*$.
Thus locally $P\ge L$.

On the other hand, notice that
  $L(a^2=0) >P(a^2=0)$ and
$L(a^2=\infty )>P(a^2=\infty )$, for all values of $R$.

 $L$ and $P$ are differentiable functions of their arguments.
Combining the above facts  one infers that, due to the continuity
of $L$ and $P$, there must exist at least two solutions in a neighbourhood
of $ R_*$. Those solutions coincide at $R=R_*$, due to the above
construction.
The set of those points constitutes at least two   branches.
Since $\partial_{a^2}(L-P)=1- 4U^2/(p^2R^2a^2)\ne 0$ at any point of a
solution branch with $R\ne R_*$, the implicit function argument
would be used to extend the interval of the existence
onto a whole bounded domain.
Those solutions  are differentiable
for $R\ne R_*$.

One of the solutions is supersonic below $R_*$ and subsonic above $R_*$
and it can be interpreted as describing collapse of matter onto
a  black hole. The other solution is subsonic  for $R<R_*$
and supersonic above; it can correspond to an exploding gas.

\subsection{Qualitative results}

In what follows we shall deal with  a solution that is subsonic
asymptotically, i. e., describes accretion of a fluid.

{\bf Theorem 3}.    An asymptotically subsonic solution of the system
(\ref{V.1} - \ref{V.5})   satisfies following conditions:

i) If  $R\ne R_*$ then  $\partial_R(U^2R^4) >0$ and
the speed of sound decreases,
$\partial_Ra^2 \le 0$, with the equality only at spatial infinity;

ii)  $U^2 > {m\over 2R}$ for $R<R_*$ and $U^2 < {m\over 2R}$
for $R> R_*$.

iii) Inside the supersonic region $a^2(pR)^2/4< U^2\le 2m/R$.

iv)  Mass density $\rho $ monotonously decreases and $\rho $ is  bounded
in the supersonic region, $R<R_*$,
\begin{equation}
\rho \le \rho_{\infty } \bigl[ 1 +
(\Gamma -1) {4m\over Ra^2_{\infty }}\bigr]^{1\over \Gamma -1} .
\label{VI.0}
\end{equation}

The proof is postponed to the Appendix.

The estimates of ii) and iii) in Theorem 3 require an explanation.
It proved to be convenient  to define a sonic point by  requiring
that $a^2(pR)^2/4=U^2$ instead of the condition (used in the Bondi model)
$a^2=U^2$. Therefore the speed of sound  can be bigger
than infall velocity in regions close to horizons if the factor $pR/2$
is significantly smaller than 1. In the traditional terminology such a
solution would be called subsonic. Numerical data of the next
Section show that the value $|U|/a$ at a horizon  depends strongly on
the location of a sonic point -  on the ratio $R_*/(2m)$, which in turn
depends on the asymptotic speed of sound $a_{\infty }$.   $|U|/a$ decreases
with the increase of the asymptotic speed of sound.

\section{  Bondi model and the relativistic solution}

 The insertion  of  (\ref{V.3}) and
(\ref{V.5}) into  (\ref{4.18}) (with $c_*=0$)
 yields  the mass accretion  rate
\begin{equation}
\dot m =\pi m^2{\rho_{\infty }\over a^3_{\infty }}
\Bigl( {a^2_*\over a^{2}_{\infty }}\Bigr)^{(5-3\Gamma )/2(\Gamma -1)}
(1+{a^2_{* }\over \Gamma })( 1+3a_*^2).
\label{VIII.1}
\end{equation}
One can write that as
\begin{equation}
\dot m = \Omega \dot m_B,
\label{VIII.1a}
\end{equation}
where
\begin{equation}
\dot m_B=\pi m^2{\rho_{\infty }\over a^3_{\infty }}
\Bigl( {2\over 5-3\Gamma }\Bigr)^{(5-3\Gamma )/2(\Gamma -1)}
\label{VIII.1b}
\end{equation}
 is the  mass accretion rate predicted by the Bondi model and
\begin{equation}
\Omega =
\Bigl( {(5-3\Gamma )a^2_*\over 2a^{2}_{\infty }}\Bigr)^{(5-3\Gamma )/2
(\Gamma -1)}
 (1+3a^2_*)(1+{a^2_{* }\over \Gamma }).
\label{VIII.2}
\end{equation}
$\Omega $ can be interpreted as the relativistic correction factor.

Application of Theorem 2 leads to useful estimates for $\Omega $.

{\bf Theorem 4}.  Assume $1\le \Gamma \le 5/3$.
The relativistic correction factor satisfies
\begin{eqnarray}
&&(1+3a^2_*)(1+{a^2_*\over \Gamma })    \nonumber\\
&&\ge  \Omega   \ge  \nonumber\\
&&(1+3a^2_*)(1+{a^2_*\over \Gamma })e^{-C },
\label{VIII.3a}
\end{eqnarray}
where
\begin{equation}
C= {a^2_*\over 1+3a^2_*}\Bigl( (4.5- {2.5\over \Gamma })
\ln 4-{0.75\over \Gamma }\Bigr) ={a^2_*\over 1+3a^2_*}
\Bigl( 6.24-{4.22\over
 \Gamma } \Bigr) .
\label{VIII.4}
\end{equation}

{\bf Proof.}  The definition of
$\Omega $ and Theorem 2 yield  immediately the forthcoming inequalities,
\begin{eqnarray}
&& {(1+3a^2_*)(1+{a^2_*\over \Gamma })\over \Bigl( 1+ {
2B\over  (5-3\Gamma )}{3a^2_*\over 1+3a^2_*}
\Bigr)^{(5-3\Gamma )/2(\Gamma -1)}} \ge   \nonumber\\
&& \Omega   \ge  \nonumber\\
&&{(1+3a^2_*)(1+{a^2_*\over \Gamma }) \over
 \bigl( 1+
{2(\Gamma -1)\over 5-3\Gamma }C\bigr)^{(5-3\Gamma )/2(\Gamma -1)} }.
\label{VIII.3}
\end{eqnarray}
Bounding from above the left hand side of (\ref{VIII.3})  by
$(1+3a^2_*)(1+{a^2_*\over \Gamma })$ yields the first bound of Theorem 4.
 The proof of the other inequality bases on the obvious estimate
\begin{equation}
 \bigl( 1+
{2(\Gamma -1)\over 5-3\Gamma }C\bigr)^{(5-3\Gamma )/2(\Gamma -1)}
\le e^{C}.
\end{equation}
Taking into account  Theorem 2 and its implications stated in (\ref{c1})
and below formula (\ref{c2}),
 one can write the above in terms of
asymptotic data,
\begin{equation}
(1+3 {\delta_0-\sqrt{\delta_0^2+4a_{\infty }^2\delta_2}\over 2\delta_2} )
(1+
 {\delta_0-\sqrt{\delta_0^2+4a_{\infty }^2\delta_2}\over 2\delta_2}
 {1\over
\Gamma   })
 \ge \Omega \ge   (1+4.8a^2_{\infty })(1+{1.6a^2_{\infty }\over \Gamma })
 e^{-C}.
\end{equation}

The  relativistic correction factor $\Omega $ is close to 1 when
$a_*^2 <<1$, i. e., when the asymptotic gas temperature is low.
 $\Omega $ is bounded from below by 0.99. (\ref{VIII.3a}) yields,
in the ultrarelativistic regime $a^2_*\approx 1 $,
\begin{equation}
4(1+{1\over \Gamma })\ge \Omega \ge 1.6(1+{1\over \Gamma }).
\end{equation}
Ultrarelativistic effects  enhance accretion,  with the strongest effect
for the isothermal gas with $\Gamma =1$.
The enhancement is  smaller for $\Gamma \approx 5/3$, as seen
from  the preceding estimate.

The Bondi model fails only in
describing the hot gas  mode. The correction factor $\Omega $ tends
quickly  to 1 when sonic points are  far away from the Schwarzschild sphere,
For instance, if $\Gamma =4/3$ then  $\Omega <7$  at $R_*/m=2$ but
$\Omega <1.1$ at  $R_*/m=25$.

We analyse numerically a relativistic gas, with the adiabatic index $\Gamma
 = 4/3$, falling onto a black
hole. Results complement analytic estimates and they
 are comprised in the table.

\begin{tabular}{|c|c|c|c|c|c|c|}
\hline
 &\multicolumn{6}{c||}{ }\\
$R_*/(2m)$  & $a^2(R_*)/a_{\infty }^2$  &
$^*~a (R_n)$ & $^* ~U(R_n))$
 & $^*\rho (R_n))/\rho_{\infty }$ & $\rho (R_*)/\rho_{\infty }$
 &  $\rho (2\times R_*)/\rho_{\infty }$ \\
 \hline
\cline{2-6}
500000   &2       & 0.168 & 0.999 & 1.41$\times 10^9$  & 8    & 3.14 \\
50       & 1.99   &0.173  &0.92   & 1595               &7.89  &3.9   \\
5        &1.91    &0.34   & 0.78  & 56.4               &6.95       &3.46  \\
1.1      & 1.62   &0.86   & 0.53  &4.82                & 4.24 &2.1    \\
\cline{2-6}
\hline
\multicolumn{6}{l}{$^*$~$R_n=1.001\times 2m$} \\
\hline
\end{tabular}

Some features of accreting  solutions depend in a crucial way on the location
of sonic points. When sonic points are  close to a horizon, the speed of
sound is close to one  while the infall velocity at a horizon is smaller
and it barely exceeds  1/2. When sonic points are far away from a horizon,
$R_*>> 2m$, the infall
velocity nears to the speed of   free fall ($U\approx  1$ close
to a horizon) while the speed of sound is then much smaller than $U$.
An interesting fact is that the energy density changes quite moderately
- by a factor of the order of unity - if sonic points are close to the
Schwarzschild sphere. In contrast with that, solutions with $R_*>>2m$ are
characterized by a rapid growth - up to ten orders  -  of the energy density
near the horizon. The energy density changes by a factor not greater than 8
in the region exterior to a sonic point with $R\epsilon (R_*, \infty )$; that
type of moderate decay is common for all solutions, irrespective of
the value of $2m/R_*$. That actually follows from Theorem 2, which bounds
$a^2_*/a^2_{\infty }$ and - consequently - also $\rho (R_*)/\rho_{\infty }$.
Solutions with sonic points close to a horizon have $a$ approaching 1 and
they describe a high temperature (circa $10^{10}$ K) gas,
 with $a_{\infty }\approx 0.5$.

{\bf Acknowledgements.}  This work has been supported by
 the KBN grant 2 PO3B 010 16.

\vfill \eject

\centerline{\bf Appendix}

{\bf Proof of Theorem 3.}

1) Equation (\ref{4.14a})  yields, ignoring the backreaction term,
the crucial  relation
\begin{equation}
\partial_R(U^2R^4) \Bigl( 1-{4U^2\over a^2p^2R^2}\Bigr) =
{16R\over p^2} ({m\over 2R }- U^2).
\label{VI.1}
\end{equation}
2) The first observation, that signs of $\partial_Ra^2$ and
of $\partial_R(U^2R$)  are opposite and that they  vanish simultaneously
at finite values of $R$, can be drawn  from (\ref{4.14}).

3) Let $R_*$ be a position of the
sonic point; thus $U^2(R_*)=m/2R_*$.
Assume that in the vicinity
of $R_*$ the expression  $\partial_R(U^2R^4) $ is strictly negative.
Then (\ref{VI.1}) yields $U^2R < m/2$ for $R< R_*$ and
$U^2R > m/2$ for $R > R_*$. Therefore $U^2R$ is increasing in the region
of interest and that is incompatible with  the assumption
that $\partial_R(U^2R^4) $ is strictly negative. Thence
it must be  weakly positive  at least around the outermost sonic point.
That in turn implies that in a neighbourhood of a sonic point $2U^2R < m $
for $R>R_*$  and  $2U^2R > m $ for $R<R_*$

4) The  expression   $\partial_R(U^2R^4) $   cannot have zeroes.
For, let it vanish at some $R_1> R_*$; (\ref{VI.1}) gives
$U^2(R_1)= {m\over 2R_1}$ and we would have
$\partial_R(U^2R)\ge 0$ at $R_1$.
But that is incompatible with the assumption that
$\partial_R(U^2R^4)= 0 $ at $R_1$.

Let us now consider a region  $R<R_*$. If $R_1$ is a zero point of
$\partial_R(U^2R^4) $ but the latter does not change sign at $R_1$, then
$2U^2R$ decreases for $R<R_1$  towards the value $m$ and increases for
$R>R_1 $ (due to estimates proven in the final part of 3)),
Hence $\partial_R(U^2R)= 0$ at $R_1$.   But that
 contradicts $\partial_R(U^2R^4) =0$ at $R_1$. Similarly,
if $\partial_R(U^2R^4) $  changes sign in the vicinity of $R_1$, then
we are led to contradiction.

Thus   $\partial_R(U^2R^4)>0$ in the domain
of existence of the solution.  That  implies, with conjunction with
(\ref{VI.1}), that in the supersonic zone $U^2> m/(2R)$ and that $U^2<m/(2R)$
in the subsonic zone ($R>R_*$). That accomplishes the proof of ii)

5) Rewrite Eq. (\ref{4.13}), with backreaction terms being dropped
out,
\begin{equation}
\partial_R\ln (a^2+\Gamma ) = {-4\kappa \over p^2R^3}\Bigl( {m\over R}-
{U^2\over 2R}+ {1\over 2}\partial_R(U^2R) )\Bigr) .
\label{VI.2}
\end{equation}
Let $R$ be a largest  point $R<R_*$  such that $U^2R=2m$;
then  $\partial_R(U^2R)|_R<0$ and from (\ref{VI.2}) follows
$\partial_Ra^2|_*>0$, in contradiction with hitherto proven
monotonic falloff of $a^2$.  That shows the bounds of  iii) - that
$U^2R<2m$. (\ref{VI.1}) and ii) imply, in the supersonic region, $a^2p^2R^2/4 <1$.
Extremal values of  the speed of sound
(achieved at a horizon of a black hole)  cannot exceed $4/(pR)^2$
while the speed of infalling particles does not exceed 1, the speed of light.

6)  The decrease of the speed of sound together with   (\ref{4.17})  lead to
the conclusion that  the mass density  also  decreases,
$\partial_Ra^2\le 0$ and $\partial_R\rho \le 0$.
  The numerical estimates of iv)
are obtained from  inserting inequalities proven in 5) to the
expression (\ref{4.17}).

That ends the proof of Theorem 3.


\begin{references}
\bibitem{Michel} F. C. Michel, {\it Ap. Space Sci.} {\bf 15}, 153(1972);
\bibitem{Shapiro} S. L. Shapiro, S. A. Teukolsky, Black holes,
white dwarfs and neutron stars, John Wiley and Sons, Inc. 1983;
\bibitem{Bondi}  Bondi, H. MNRAS {\bf 112}, 195(1952);
\bibitem{Tolman} Lemaitre, G. {\it Ann. Soc. Sci. Bruxelles}
{\bf A53}, 51(1933); Tolman, R. C., Relativistic Thermodynamics and
Cosmology, Oxford: Clarendon Press 1934;
Podurets, M. A. {\it  Soviet Astron.} {\bf 8}, 19(1964); Ch. W. Misner and
D. H. Sharp, {\it Phys. Rev. } {\bf 136B}, 571(1964);
\bibitem{Novikov} I. D. Novikov, K. S. Thorne,
{\it Black Hole Astrophysics} in Black Holes, C. deWitt and B. deWitt,
editors, Gordon and Breach, New York 1973;
\bibitem{hawking}    S. W. Hawking and G. F. Ellis, The large scale
structure of spacetime, Cambridge University Press 1973;
\bibitem{Rosz} K. Roszkowski, private  communication.
\end{references}
\end{document}